\documentclass[12pt]{article}
\usepackage[cp1251]{inputenc}
\usepackage[english]{babel}
\usepackage{amssymb,amsmath,textcomp}
\usepackage{graphicx}
\begin{document}
\newcommand{\dd}[2]{\frac{\partial #1}{\partial #2}}

 \begin{center}
 \section*{Photon gas in a finite box:\\thermodynamics and finite size effects}
 
 A.~A.~Sokolsky and M.~A.~Gorlach
 \par \medskip {\it Physics Department of the Belarusian State University, 4 Nezalejnosty av.,\\Minsk, 220030, Belarus.}
  
 \par \vspace{14 pt} {\footnotesize Thermodynamic properties of a photon gas in a small box are explored taking into account\\finite size effects. General thermodynamic relations are derived for this finite system.\\Photon gas thermodynamic functions are calculated for the case of cuboid cavity\\on the basis of first principles. New finite size effects are discussed.}
 \end{center}

 \par\subsection*{Introduction}
 \par \hangindent=0.7 cm \hangafter=-1 As it is known, the phenomenon of blackbody radiation plays an important role in physics. In 1884 L.~Boltzmann derived that internal energy of radiation in a cavity is proportional to the fourth power of the temperature; this statement is known as Stefan-Boltzmann law~(\cite{Boltzmann}). In 1900, basing of quantum hypothesis, M.~Planck obtained a formula describing blackbody radiation spectrum~(\cite{Planck}); this formula was thought to be correct for arbitrary temperatures and cavity volumes for a long period of time.
 \par \hangafter=-1 But it turns out that Planck's formula, Stefan-Boltzmann law, Wien's displacement law are not accurate in the case of sufficiently small cavities and low temperatures. This remark was apparently first done by Bijl~(\cite{Bijl}). He also found a criterion of the Planck's formula validity: 
 
 \begin{equation}\label{crit}
 T\,V^{1/3}\gg B\equiv \frac{\hbar c}{k_B}\approx 0.2290\,cm\cdot K
 \end{equation}
 
 \par \hangafter=-1 The following step was done in~\cite{Case} where first correction terms to the Stefan-Boltzmann law were calculated for the case of cubic cavity with ideally conducting walls. Since that time numerous efforts were made in order to take into account finite size effects in thermal radiation theory. The list of main research activities in this field, far from being complete, is given below.
\begin{itemize}
  \item	Weyl's problem~(\cite{Weyl}): calculation of the eigenvalues distribution of vector wave equation in cavities of various shapes, averaged expressions for electromagnetic mode density, etc. (\cite{Baltes1972}, \cite{Baltes1976}, \cite{Hilf});
  \item Refinements of the Stefan-Boltzmann formula and corresponding expressions for entropy, specific heat and radiation pressure, study of cavities having various shapes with ideally conducting walls~(\cite{Baltes1972}, \cite{Baltes1973}, \cite{Hilf});
  \item Corrections to the Stefan-Boltzmann law due to finite conductivity of cavity walls~(\cite{Eckhardt});
  \item Consideration of radiation from small particles, exploration of a particle size influence on thermal radiation spectrum~(\cite{McGregor}, \cite{Reiser}).
\end{itemize} 
 
Interest to the finite size effects continues till present time. Calculation of thermodynamic functions of a hot quark-gluon plasma was performed in \cite{Gliozzi} with consideration of finite size effects. In \cite{Garcia} it was shown that the experimental detection of deviations from the Planck's formula is within the reach of current experimental capabilities. Finally, in \cite{Reiser} the experimental observation of such deviations was reported: it was detected that in narrow spectral range thermal radiation may exceed the value predicted by the Planck's formula. 
\par \hangafter=-1 That is why it is of particular interest to study finite size effects for equilibrium thermal radiation. In present paper we develop thermodynamic  approach to the above mentioned problem, perform calculations of thermodynamic functions of electromagnetic radiation in a cavity having the form of rectangular parallelepiped (cuboid), and discuss new finite size effects.

 \par\subsection*{General thermodynamic relations for the photon gas in a finite box} 
 \par\hangindent=0.7 cm\hangafter=-1 Thermodynamics of small systems deals with average values of physical quantities. These average values may be comparable with fluctuations for small systems; however, in some cases averages are of interest. Thermodynamics of small systems was discussed in the literature in various contexts (see for example \cite{Hilf}, \cite{Hill}), but it is relevant to mention the case of the photon gas separately.
 \par \hangafter=-1 Let us consider electromagnetic radiation in a finite cavity of arbitrary shape. It is well known that electromagnetic field in the cavity may be represented from mathematical point of view as an ensemble of noninteracting harmonic oscillators (modes) corresponding to the eigenfrequencies of the cavity, the total field energy being a sum of oscillators energies: $E=\sum\limits_m E_m$, where $m$ enumerates oscillators of the ensemble  (\cite{landau2}). As oscillators are noninteracting they are statistically independent. Note that oscillators in consideration are distinguishable as they have either different frequencies or correspond to different wavevector directions or to different mode polarizations. That is why the Gibbs distribution is applicable to the single oscillator

 \begin{equation}\label{gibbs}
 w_m\left(r\right)=\mathrm{exp}\left(\frac{F_m-E_m\left(r\right)}{k_B\,T}\right) 
 \end{equation}
 r enumerates states of $m^{\mathrm{th}}$ oscillator. Relation (\ref{gibbs}) is valid for radiation in small cavities at low temperatures as well as for ``standard" case (corresponding to satisfaction of the Bijl's criterion (\ref{crit})).
 \par \hangafter=-1 Applying canonical distribution method to the case under consideration and using quantum-mechanical expression for the harmonic oscillator energy levels, we obtain an equation for radiation free energy $F\equiv\sum\limits_m F_m$:
 
 \begin{equation}\label{freeenergy1}
 F=F_{vac}+k_B\,T\,\sum\limits_m \mathrm{ln}\left(1-\mathrm{exp}\left(-\frac{\hbar\,\omega_m}{k_B\,T}\right)\right)
 \end{equation} 
Summation over all modes of the cavity is implied in (\ref{freeenergy1}). $F_{vac}=\frac{1}{2}\,\sum\limits_m \hbar\,\omega_m$~--- vacuum energy needed in order to consider vacuum effects (like Casimir effect). We will omit the $F_{vac}$ term in the following consideration concerning only with radiation energy (as the result, Casimir forces will not be taken into account in the expression for $p$ ).
\par \hangafter=-1 Eigenfrequencies of the cavity $\omega_m$ depend not only on the volume of this cavity but also on its shape. So, one may expect that thermodynamic functions of radiation may depend on the cavity shape in obvious contrast with ``standard" thermodynamics. We illustrate this conclusion by the results of calculations below. The derivation of thermodynamic relations for general (``standard") case from the Gibbs distribution is discussed in \cite{landau5}. Taking into account the pecularity of our particular case, we arrive to the relation of the form
 
 \begin{equation}\label{1ntd}
 dE=T\,dS-p\,dV+\sum\limits_i \Lambda_i\,d\lambda_i
 \end{equation}
$E$ is average radiation energy ($E=-T^2\,\frac{\partial}{\partial\,T}\left(\frac{F}{T}\right)_{V,\lambda_i}$), $S=\frac{E-F}{T}$ is radiation entropy, pressure is defined as $p\equiv -\left(\dd{E}{V}\right)_{S,\lambda_i}=-\left(\dd{F}{V}\right)_{T,\lambda_i}$ ,
 $\Lambda_i=\left(\dd{E}{\lambda_i}\right)_{S,V,..}=\left(\dd{F}{\lambda_i}\right)_{T,V,..}$.
  $\lambda_i$~are interpreted as dimensionless parameters defining the cavity shape. For example, only two parameters are needed in order to describe a cuboid shape. Definition of pressure given above and the usual one are alike; but the derivative is calculated at a fixed cavity shape in first definition and $p$ may be considered as pressure averaged in certain way. The relation between this ``pressure" and the electromagnetic forces acting on a cavity walls may be not so simple as in ``standard" case. 
 \par\hangafter=-1 From similarity considerations it's helpful to denote $\omega_n=\tilde{\omega}_nc\,V^{-1/3}$ with the dimensionless parameter $\tilde{\omega}_n$ independent on the cavity volume. Then (\ref{freeenergy1}) may be rewritten as
 
 \begin{equation}\label{freeenergy2}
 F=k_B\,T\,\sum\limits_m \mathrm{ln}\left(1-\mathrm{exp}\left(-\frac{\hbar\,c\,\tilde{\omega}_m}{k_B\,T\,V^{1/3}}\right)\right)
 \end{equation}
 
 Thus, the expression for the radiation free energy in the cavity of arbitrary shape is as follows:
 \begin{equation}\label{freeenergy3}
 F=T\,f\left(T\,V^{1/3}\right)
 \end{equation}
 $f$ is a shape dependent function. Now we may determine radiation energy and pressure from equation (\ref{freeenergy3}). The comparison of the obtained expressions leads to a conclusion
 \begin{equation}\label{energy1}
 E=3p\,V
 \end{equation}
Consequently, the equation of state (\ref{energy1}) well known in ``standard" case is also correct for the radiation in a finite cavity (with used definition of  pressure). The radiation energy for the cavity of arbitrary shape obtained from (\ref{freeenergy3}) is 
 \begin{equation}\label{energy2}
 E=\frac{4\sigma}{c}\,V\,T^4\,\varphi(T\,V^{1/3})
 \end{equation}
$\sigma$ is a Stefan-Boltzmann constant, $\varphi(x)=-\frac{c}{4\sigma}\,\frac{f'(x)}{x^2}$; comparison with well-known ``standard" expression gives $\lim\limits_{x \to\infty}\varphi(x)=1$.  
 \par\hangafter=-1 The explicit expressions for function $\varphi$ were deduced in particular cases in \cite{Case}, \cite{Hilf}, etc. It must be noted that these analytical expressions have an asymptotic character, i.e. they are valid either under the condition $T\,V^{1/3}\gg B$ (so-called high-temperature expansion) or under the condition $T\,V^{1/3}\ll B$ (so-called low-temperature expansion). Description of the intermediate temperature region requires numerical computation (see Fig.~\ref{ris:Picture1} below).

\par\subsection*{Computation technique}
\par\hangindent=0.7 cm\hangafter=-1 Basing of general results discussed in the previous section, we consider the particular case of cuboid cavity with edges $X$, $Y$, $Z$. The shape of such a cavity is described by two dimensionless parameters $\alpha=\frac{X}{Z}$, $\beta=\frac{Y}{Z}$; let's designate $a=V^{1/3}=\left(X\,Y\,Z\right)^{1/3}$. Then normalized eigenfrequencies of such a cavity are
 \begin{equation}\label{freq}
\tilde{\omega}_n \equiv \frac{\omega_n\,a}{c}=\frac{\pi}{(\alpha\,\beta)^{2/3}}\,\sqrt{n^2_x\,\beta^2+n^2_y\,\alpha^2+n^2_z\,\alpha^2\,\beta^2}
 \end{equation}
$n=\{n_x,n_y,n_z\}$; the mode exists if two or three numbers from set $n$ are nonzero; in the first case degeneracy is $g_n=1$, in the second --- $g_n=2$ (additional degeneracy may appear due to the symmetry of the cavity) (\cite{landau8}). We use direct summation in (\ref{freeenergy2}) while $\tilde{\omega}\leqslant\tilde{\omega}_e$, for larger frequencies the rest of the sum is replaced by the corresponding integral. Final result depends to some extent on $\tilde{\omega}_e$ choice. This unwanted circumstance may be removed by $\tilde{\omega}_e$ variation, so final result becomes insensible to the $\tilde{\omega}_e$ increase. The calculation formula is 

 \begin{equation}\label{energy3}
 \begin{split}
 \frac{E}{k_B\,T}&=\sum\limits_{\substack{n,  \\{\tilde{\omega}_n}\leqslant\tilde{\omega}_e}}    g_n\,\frac{B\,\tilde{\omega}_n}{Ta}\,\left(\mathrm{exp}\left(\frac{B\,\tilde{\omega}_n}{Ta}\right)-1\right)^{-1}+\\
 &+\int\limits_{\tilde{\omega}_e}^{\infty} \frac{B\,\tilde{\omega}^3}{\pi^2\,Ta}\,\left(\mathrm{exp}\left(\frac{B\,\tilde{\omega}}{Ta}\right)-1\right)^{-1}\,d\tilde{\omega}
 \end{split}
 \end{equation}
$B=\frac{\hbar\,c}{k_B}\approx\,0.2290\,cm\cdot K$, $n$ enumerates different eigenfrequencies. Inclusion of the integral term to the formula~(\ref{energy3}) reduces the calculation time. Another thermodynamic functions were calculated similarly. Some of the obtained results are presented below. Fig.~\ref{ris:Picture1} illustrates the domain of validity of the asymptotic formulae \cite{Baltes1973} for the radiation energy. Comprehensive discussion of this question for cubic cavity is given in \cite{Baltes1973}. The influence of the cavity shape on thermodynamic functions of the photon gas is illustrated on Fig.~\ref{ris:Picture2}.
  
    \begin{figure}[h]
    \begin{center}
    \begin{minipage}[h]{0.46\linewidth}
    \includegraphics[width=1\linewidth]{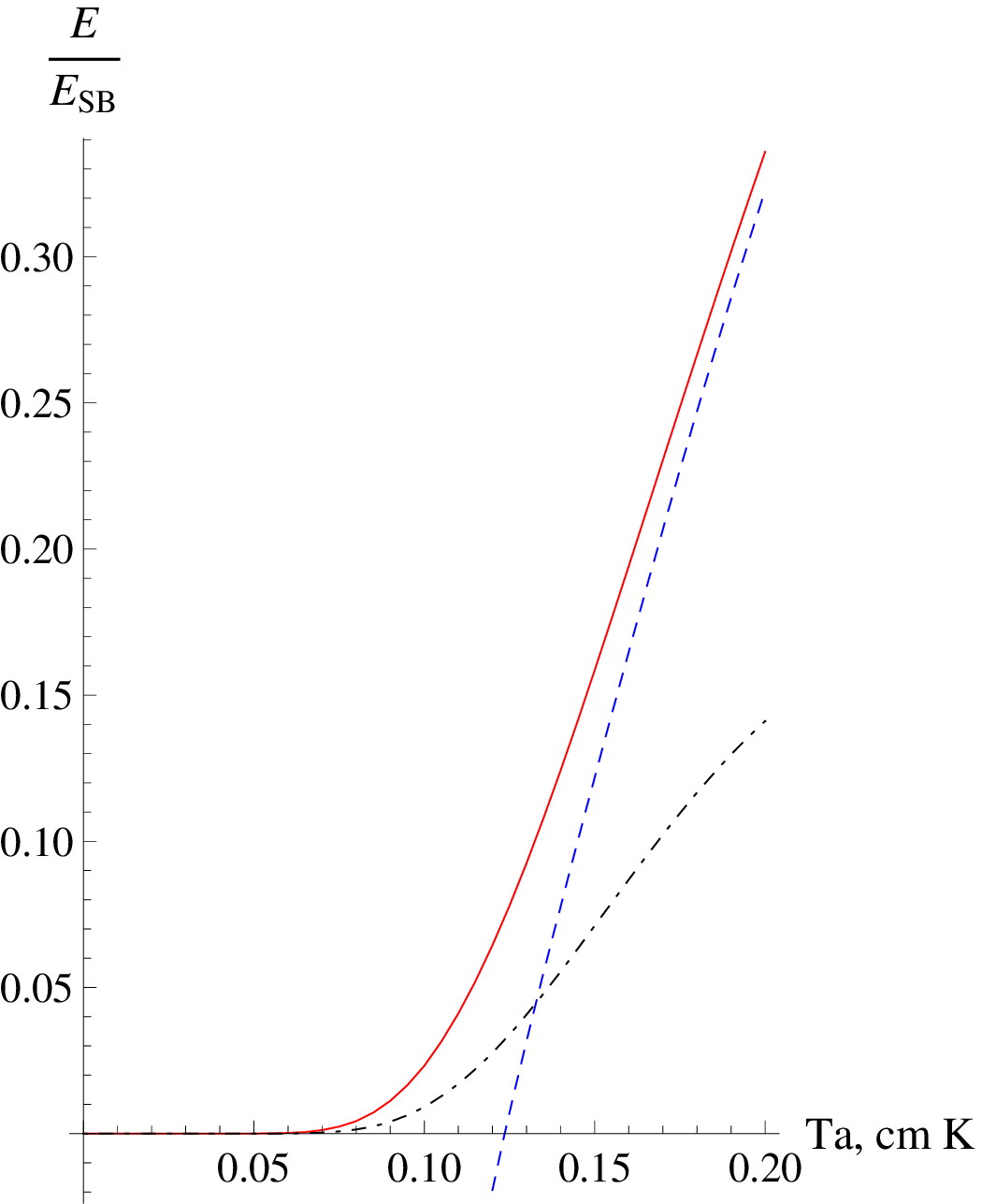}
    \caption{Radiation energy in a cubic cavity normalized to the value predicted by the Stefan-Boltzmann law: {\footnotesize solid curve~---calculated results; dashed curve~--- high-temperature expansion \cite{Baltes1973}; dot-dashed curve~--- low-temperature expansion \cite{Baltes1973}.}}
    \label{ris:Picture1}
    \end{minipage}
  \hfill 
    \begin{minipage}[h]{0.46\linewidth}
    \includegraphics[width=1\linewidth]{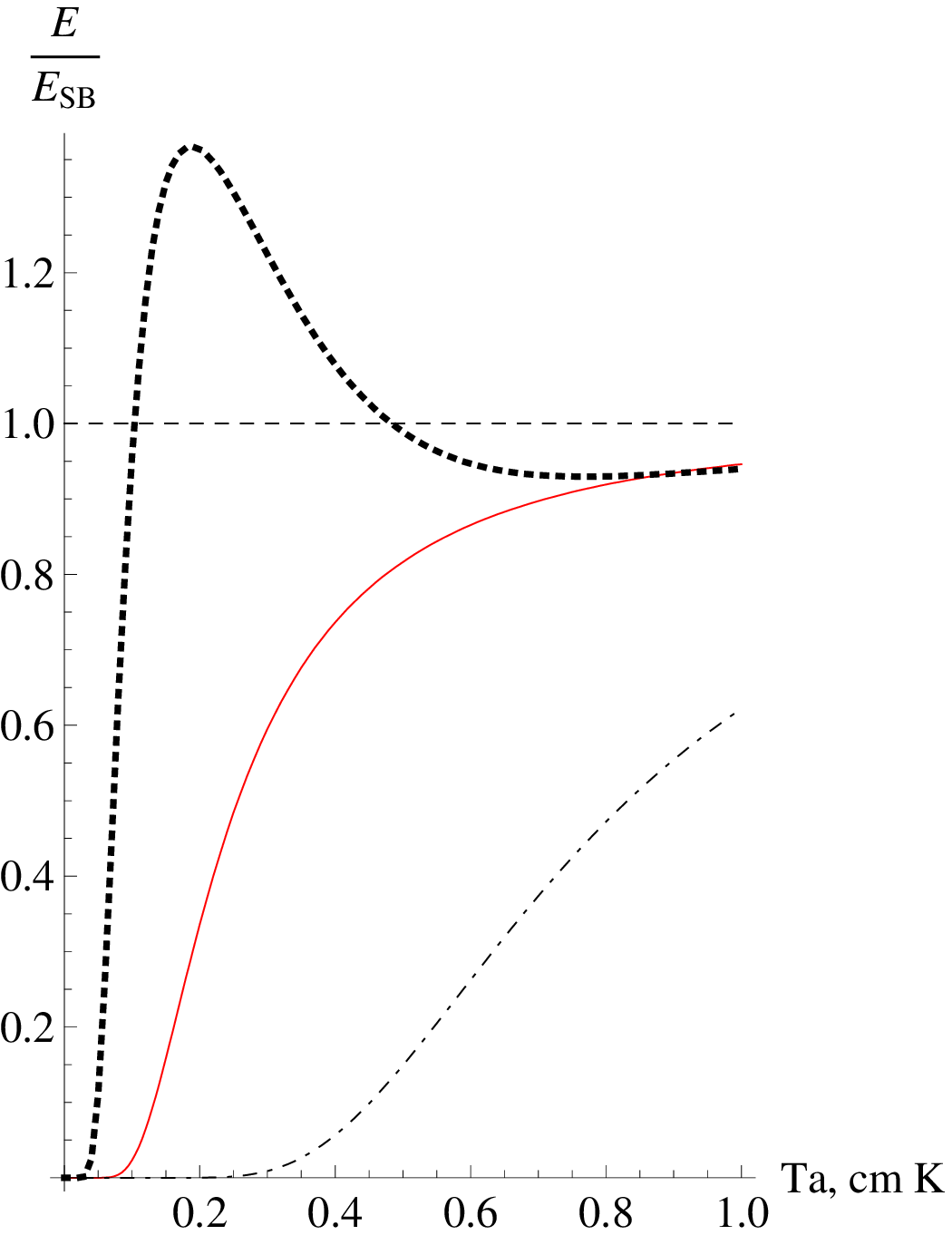}
    \caption{Radiation energy normalized to the value predicted by the Stefan-Boltzmann law for cuboid cavities of different shapes:
    \footnotesize{solid curve~--- $\alpha=\beta=1$;
    dotted curve~--- $\alpha=\beta=10$;
    dot-dashed curve~--- $\alpha=\beta=10^{-2}$.}}
    \label{ris:Picture2}
    \end{minipage}
    \end{center}
    \end{figure}

 \par\subsection*{Finite size effects in connected cavities and radiation pressure anisotropy}
\par
\begin{figure}[ht]
    \begin{center}
    \begin{minipage}[h]{0.46\linewidth}
    \includegraphics[width=1\linewidth]{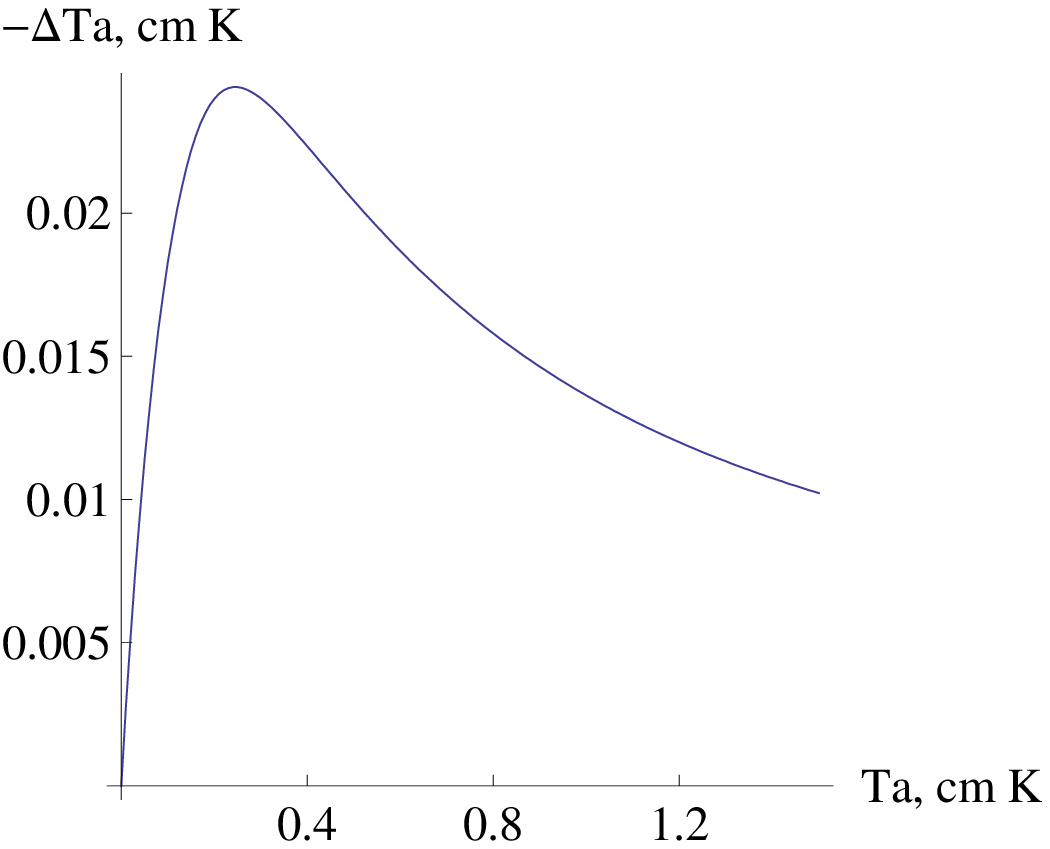}
    \caption{The temperature decrease of the photon gas in a 50 cubes when the partitions between these cubes are removed adiabatically.}
    \label{ris:Picture3}
    \end{minipage}
    \hfill 
    \begin{minipage}[h]{0.46\linewidth}
    \includegraphics[width=1\linewidth]{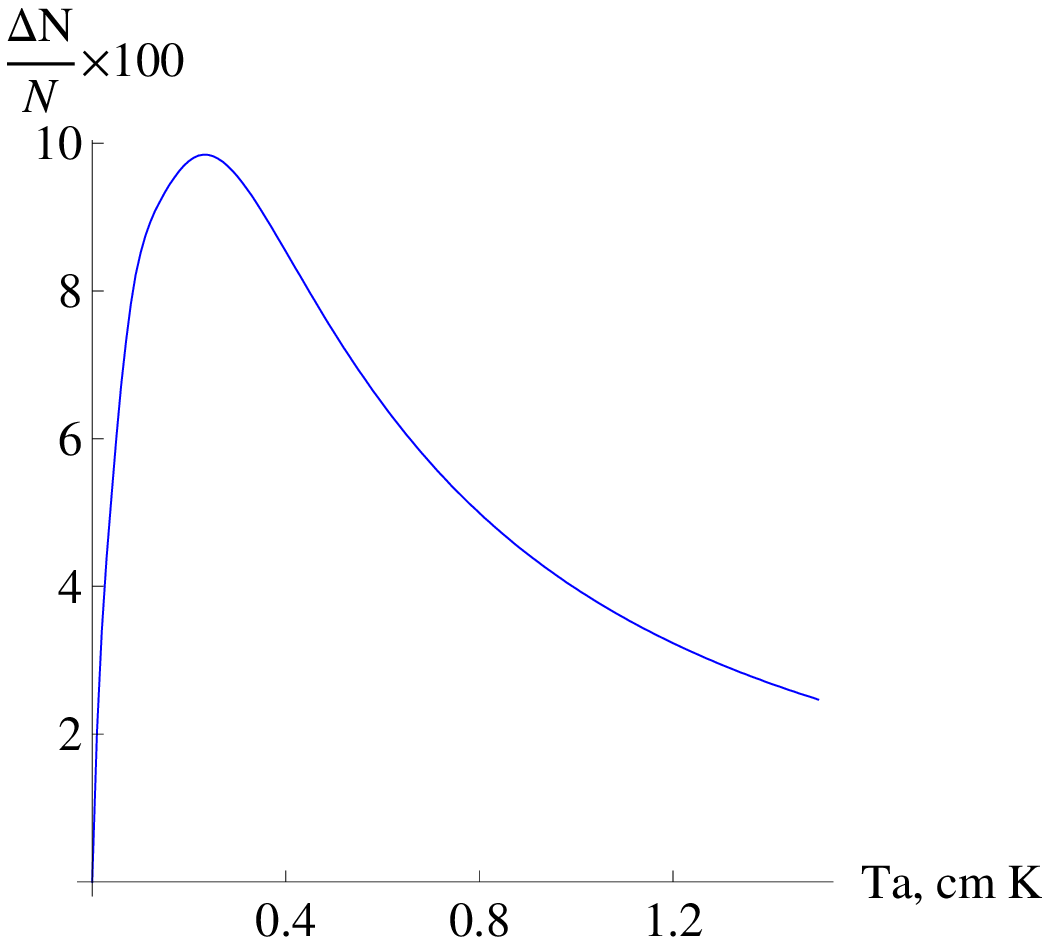}
    \caption{Relative increase of an average photon number in a system of 50 cubes when the partitions between these cubes are removed adiabatically.}
    \label{ris:Picture4}
    \end{minipage}
    \end{center}
    \end{figure}

\begin{figure}[h!]
    \begin{center}
    \begin{minipage}[h]{0.46\linewidth}
    \includegraphics[width=1\linewidth]{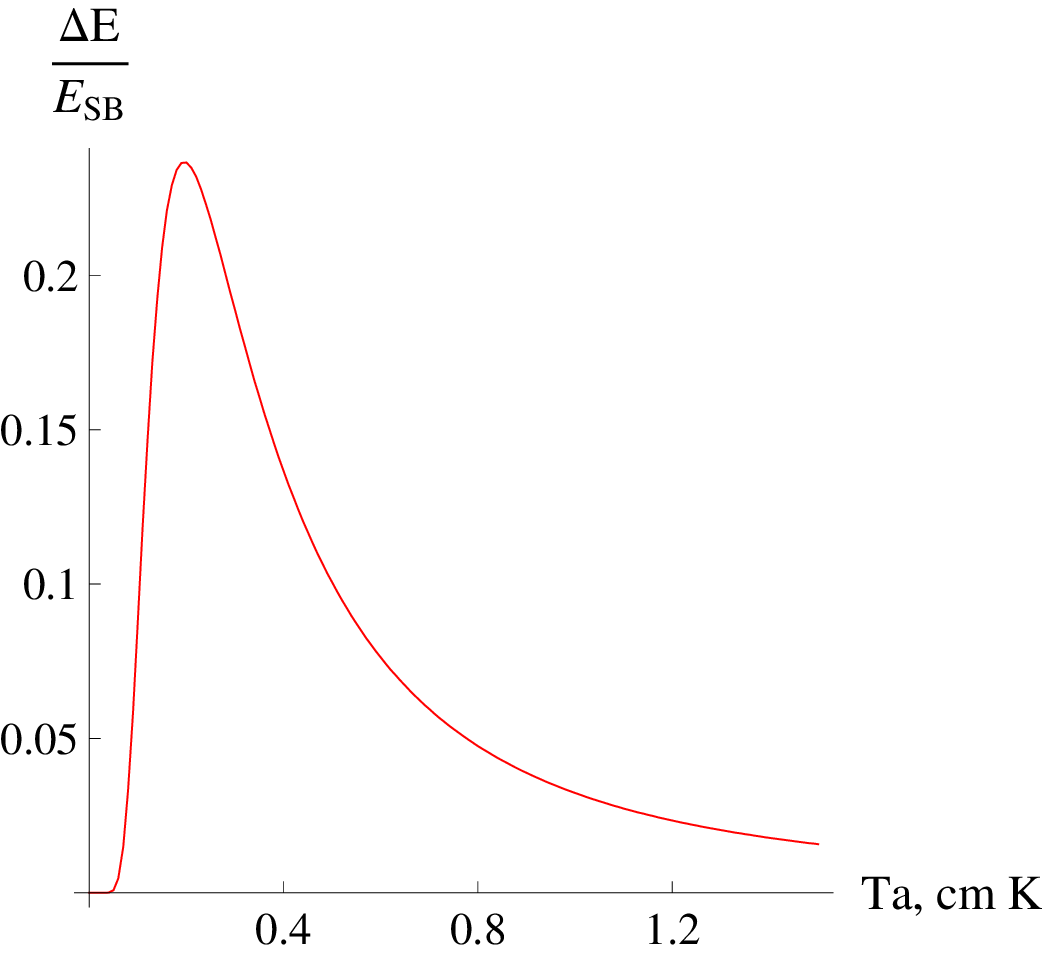}
    \caption{Extra energy supply (normalized to the total energy, calculated by the Stefan-Boltzmann law) needed when the partitions between 50 cubes are removed isothermally.}
    \label{ris:Picture5}
    \end{minipage}
    \hfill 
    \begin{minipage}[h]{0.46\linewidth}
    \includegraphics[width=1\linewidth]{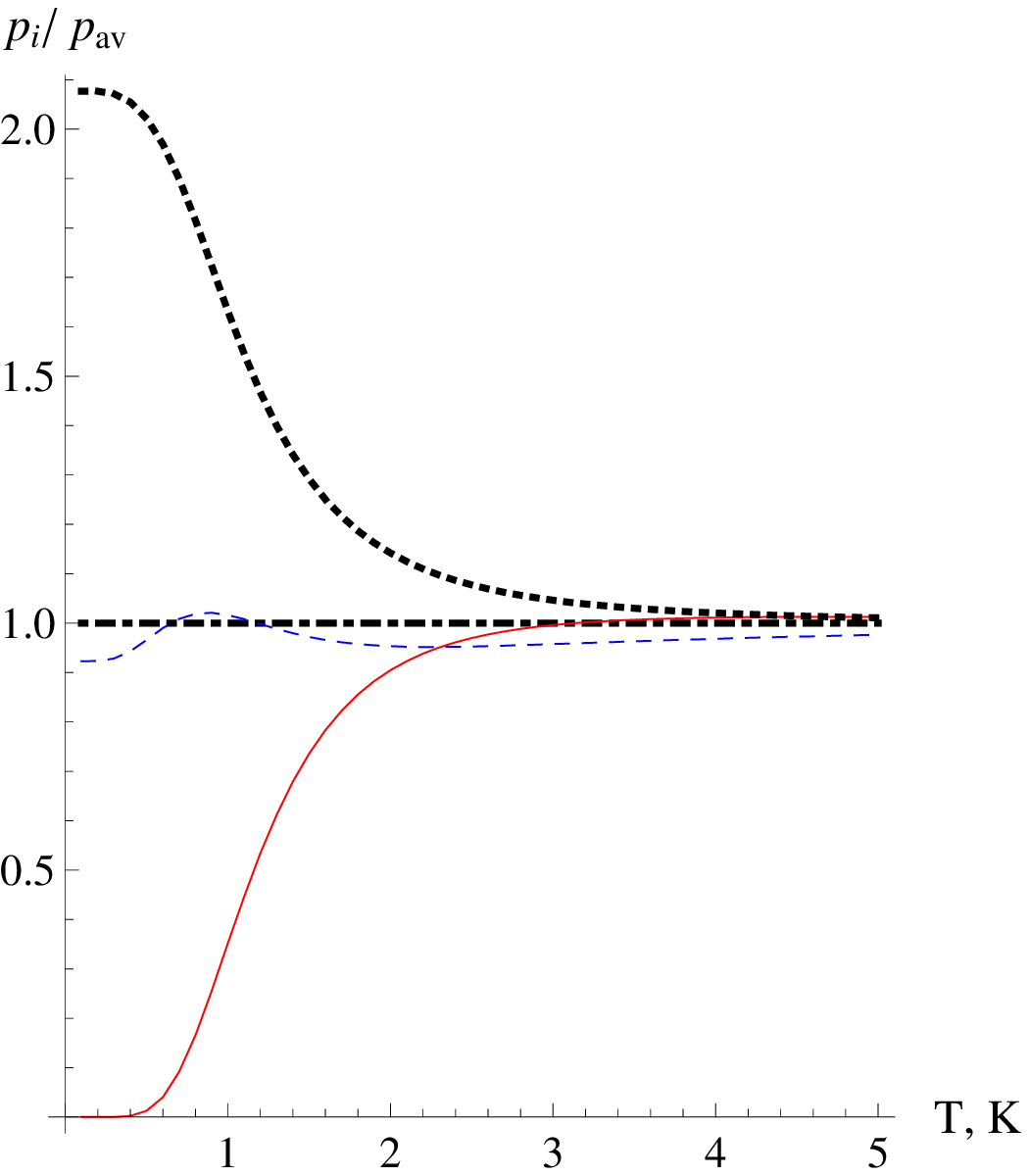}
    \caption{Pressures on the cavity faces normalized to the ``average" pressure (in (\ref{energy1})) for the cuboid cavity with edges $X=1\,\mathrm{mm}$, $Y=2\,\mathrm{mm}$, $Z=3\,\mathrm{mm}$:
    {\footnotesize solid curve~--- $p_x/p_{av}$; dotted curve~--- $p_y/p_{av}$; dashed curve~--- $p_z/p_{av}$.}}
    \label{ris:Picture6}
    \end{minipage}
    \end{center}
    \end{figure}

\hangindent=0.7 cm\hangafter=-1 Thus, shape influences considerably on the amount of radiation energy for the cavity at a fixed volume and temperature (such as $T\,V^{1/3}\sim B$). Nonlinear dependence of the energy stored in a cavity on the volume of the latter is another striking difference between ``standard" thermodynamics and thermodynamics of the photon gas in a finite box (see (\ref{energy2})). In order to clarify this statement, let's consider two identical cubic cavities separated by a partition. Then imagine that the partition is removed and now we have one cavity of doubled length and volume. If temperatures and pressures of photon gases are equal in both cavities then nothing will occur in the system from the ``standard" point of view. It turns out that the behaviour of the photon gas differs in the intermediate temperature region specified above: if the partition is removed adiabatically then the temperature of the photon gas decreases and the total number of photons increases; if the partition is removed isothermally then extra energy supply is needed. The results of calculations illustrating the effect are plotted on Fig.~\ref{ris:Picture3}, \ref{ris:Picture4}, \ref{ris:Picture5}. Calculations are performed for the case when 50 identical cubic cavities are connected. Here $a=V^{1/3}$, $V$ being the volume of a single cube.

 \par\hangafter=-1 The qualitative explanation of the effect is the following. When the partitions between cubic cavities are removed, new eigenfrequencies appear for the electromagnetic field inside. If the temperature of the system had been kept constant then the statistical weight of a given system state has increased (when new modes appear the number of ways to distribute the energy between these modes increases). Thus, the entropy has increased. On the other hand, if the entropy of the system remains unchanged (we consider the adiabatic process) the radiation temperature must decrease in accordance with Le Chatelier's principle.
	It's also desirable to explain how the radiation temperature decrease correlates with the increase of the photon number in the cavity. Note that photons corresponding to sufficiently high frequencies will be partially absorbed by walls due to the temperature decrease; but a number of photons with smaller frequencies will be also emitted by the walls in order to fill the vacancies provided by the new modes of the composite cavity. So, it turns out that the number of emitted photons is greater than the number of absorbed photons and these considerations explain the obtained results qualitatively. It should be focused that in the ``standard" case the number of photons in a cavity is proportional to the radiation entropy, the proportionality factor being independent on the cavity shape. Thus, it is obvious that the discussed effect is a typical finite size effect.
 \par\hangafter=-1 Another finite size effect for the cuboid cavity with unequal edges is radiation pressure anisotropy. That is to say, pressures on cavity faces defined as $p_x=-\frac{1}{YZ}\,\left(\dd{F}{X}\right)_{T,Y,Z}$, $p_y=-\frac{1}{XZ}\,\left(\dd{F}{Y}\right)_{T,X,Z}$, $p_z=-\frac{1}{XY}\,\left(\dd{F}{Z}\right)_{T,X,Y}$ may be unequal. According to (\ref{freeenergy2}) and (\ref{freq})
 
 \begin{equation}\label{pressure}
 p_x=\frac{\pi\,\hbar\,c}{V}\,\sum\limits_n \frac{g_n \left(\frac{n_x}{X}\right)^2}{\left(\mathrm{exp}\left(\frac{\hbar\,\omega_n}{k_B\,T}\right)-1\right)\,\sqrt{\left(\frac{n_x}{X}\right)^2+\left(\frac{n_y}{Y}\right)^2+\left(\frac{n_z}{Z}\right)^2}}
 \end{equation} 	

 \par\hangafter=-1 The similar expressions are valid for $p_y$ and $p_z$. It may be noticed that $p_x+p_y+p_z=\frac{E}{V}$; this relation follows also from the fact that  energy-momentum tensor trace equals zero for the case of free electromagnetic field. Numerical calculation of quantities $p_x$, $p_y$, $p_z$ shows that  pressures $p_x$, $p_y$ and $p_z$ are unequal in the region of sufficiently low temperatures, thowgh these quantities coinside in the region $T\,V^{1/3}\gg B$ (Fig~\ref{ris:Picture6}).

	\par\subsection*{Discussion and conclusions}
\par\hangindent=0.7 cm\hangafter=-1 We have studied some specific thermodynamic properties of the photon gas in a limited volume in present work. An attention is drawn to thermodynamic calculations. General expression for the photon gas internal energy is derived for the cavity of arbitrary shape from similarity considerations. Calculations are performed and new finite size effects are discussed for the particular case of cuboid cavity. It is shown that if the partition between two adjacent identical cuboid cavities with the photon gases at the same temperatures and pressures is adiabatically removed, the temperature of radiation decreases in the obtained composite cavity though the total number of photons increases. Qualitative explanation of the predicted effect is proposed. The anisitropy of radiation pressure is predicted for the case of cuboid cavity in low-temperature region.


\begin{thebibliography}{0}
 \bibitem{Boltzmann}
    L.~Boltzmann. Ableitung des Stefan'schen Gesetzes, bettreffend die Abh\"angigkeit der W\"armestrahlung von der Temperatur aus der electromagnetischen Lichttheorie.// Annalen der Physik. Vol.~258. \textnumero~6. (1884) S.~291-294.
 \bibitem{Planck}
    M.~Planck. Zur Theorie des Gesetzes der Energieverteilung im Normalspektrum.// Verhandlungen der Deutschen Physikalisch Gesellschaft. Vol.~2. \textnumero~17. (1900). S.~237-~245.
 \bibitem{Bijl}
    D.~Bijl. Note on thermal radiation at low temperatures.// Philosophical Magazine. Vol.~43. \textnumero~347 (1952). pp.~1342-1344.
 \bibitem{Case}
    K.~M.~Case, S.~C.~Chiu. Electromagnetic fluctuations in a cavity.// Phys.~Rev.~A. Vol.~1. \textnumero~4 (1970). pp.~1170-1174.
 \bibitem{Weyl}
    H.~Weyl. Das asymptotische Verteilungsgesetz der Eigenwerte linearer partieller Differentialgleichungen.// Mathematische Annalen. Vol.~71 (1912). pp.~441-479.
 \bibitem{Baltes1972}
    H.~P.~Baltes. Thermal radiation in finite cavities.// Helvetica~Physica~Acta. Vol.~45. \textnumero~3 (1972). pp.~481-529.
 \bibitem{Baltes1973}
    H.~P.~Baltes. Deviations from the Stefan-Boltzmann law at low temperatures.// Applied~Physics. Vol.~1. \textnumero~1 (1973). pp.~39-43.
 \bibitem{Baltes1976}
    H.~P.~Baltes. Planck's radiation law for finite cavities and related problems.// Infrared~Physics. Vol.~16. \textnumero~1 (1976). pp.~1-8.
 \bibitem{Hilf}
    H.~P.~Baltes, E.~R.~Hilf. Spectra of finite systems. Bibliographisches Institut. Manheim. 1976.
 \bibitem{Eckhardt}
    W.~Eckhardt. Corrections to the Stefan-Boltzmann radiation law in cavities with walls of finite conductivity.// Optics Communications. Vol.~14. \textnumero~1 (1975). pp.~95-98.
 \bibitem{McGregor}
    W.~K.~McGregor. On the radiation from small particles.// Journal of quantitative spectroscopy and radiative transfer. Vol.~19. \textnumero~6 (1978). pp.~659-664.
 \bibitem{Reiser}
    A.~Reiser, L.~Sch\"achter. Geometric effects on blackbody radiation.// Phys.~Rev.~A. Vol.~87. \textnumero~3 (2013). 033801.
 \bibitem{Gliozzi}
    F.~Gliozzi. The Stefan-Boltzmann law in a small box and the pressure deficit in hot SU(N) lattice gauge theory.// Journal~of Physics~A: Mathematical and Theoretical. Vol.~40. \textnumero~19 (2007). pp.~375-381.
 \bibitem{Garcia}
    A.~M.~Garcia-Garcia. Finite size corrections to the blackbody radiation laws.// Phys.~Rev.~A. Vol.~78. \textnumero~2 (2008). 023806.
 \bibitem{Hill}
    T.~L.~Hill. Thermodynamics of small systems.// The Journal~of Chemical Physics. Vol.~36. \textnumero~12 (1962). pp.~3182-3197.
 \bibitem{landau2}
    L.~D.~Landau, E.~M.~Lifshitz. The classical theory~of fields. $4^{\mathrm{th}}$ ed. Butterworth-Heinemann. 1975.
 \bibitem{landau5}
    L.~D.~Landau, E.~M.~Lifshitz. Statistical physics. Part 1. $3^{\mathrm{rd}}$ ed. Pergamon Press. 1980.
 \bibitem{landau8}
    L.~D.~Landau, E.~M.~Lifshitz. Electrodynamics~of continuous media. $2^{\mathrm{nd}}$ ed. Pergamon Press. 1984.  
 \end{thebibliography}
\end{document}